\documentclass[%
 preprint,
 amsmath,amssymb,
 aps,
floatfix,
]{revtex4-2}

\usepackage{graphicx}
\usepackage{lipsum}
\usepackage{dcolumn}
\usepackage{natbib}
\usepackage{multirow}
\usepackage{booktabs}
\usepackage{hyperref}
\usepackage{siunitx}
\usepackage{xcolor}
\usepackage{bm}

\usepackage{ulem}

\begin{document}

\preprint{NITEP 241}

\title{Quantification of the evaporation process during fragmentation of space-relevant nuclei on elemental targets}

\author{Sukhendu De$^1$}
\email{sukhendu\_d@ph.iitr.ac.in} 
\author{V. Choudhary$^2$}
\email{vishal.choudhary@jecrcu.edu.in}
\author{R. Chatterjee$^1$}%
\email{rchatterjee@ph.iitr.ac.in} 
\affiliation{%
  $^1$Indian Institute of Technology Roorkee, Roorkee, 247667, Uttarakhand, India
}%
\affiliation{%
$^2$Department of Physics, School of Science, JECRC University, Jaipur, 303905, Rajasthan, India}%
\author{W. Horiuchi$^{3,4,5,6}$}%
\email{whoriuchi@omu.ac.jp}
\affiliation{%
$^3$ Department of Physics, Osaka Metropolitan University, Osaka 558-8585, Japan}
\affiliation{%
$^4$ Nambu Yoichiro Institute of Theoretical and Experimental Physics (NITEP) , Osaka Metropolitan University, Osaka 558-8585, Japan}
\affiliation{%
$^5$ RIKEN Nishina Center, Wako 351-0198, Japan}
\affiliation{%
$^6$  Department of Physics, Hokkaido University, Sapporo 060-0810, Japan}
\date{\today}

\begin{abstract}
This study examines charge-changing cross sections for $^{12}\text{C}$, $^{14}\text{N}$, $^{16}\text{O}$, and $^{20}\text{Ne}$ projectiles on elemental targets (C, Al, Cu) at a beam energy of around 290 MeV/nucleon. The two-stage abrasion-ablation model is used, with the abrasion stage described via the Glauber model, incorporating validated single-nucleon density distributions from proton elastic scattering data. 
In the ablation stage, where particle evaporation occurs, the contribution to charge-changing cross sections is estimated using two approaches. First, a statistical decay model is employed to analyze the evaporation of protons following neutron removal in the abrasion stage. The second approach estimates evaporation contributions by subtracting the direct process component (abrasion) from the experimental charge-changing cross section data. A comparison between these estimated contributions from the experimental data and the predictions of the statistical model enables a systematic evaluation of the evaporation process. The key factors influencing evaporation, such as excitation energy distribution parameters, decay width of the emitted particles, and nucleon separation energies, are analyzed. A strong correlation is observed between the maximum excitation energy available for evaporation and the neutron separation energy of the projectiles across different targets, highlighting the role of evaporation dynamics in the charge-changing cross sections.

\end{abstract}


\maketitle
\section{\label{sec:level1}Introduction: Nuclear fragmentation}

The space radiation or Galactic cosmic ray (GCR) environment is a complex mixture of photons, electrons, protons, and other heavy ions with energy ranging from a few {eV} to several {TeV} per nucleon. GCR is emitted by sources beyond the solar system, principally supernovae, gamma-ray bursts, and other high-energy cosmic phenomena. As these intense particles move across space, they contact interstellar material and magnetic fields, which can trigger fragmentation and nuclear spallation~\cite{wil84}, resulting in secondary particle creation. When high-energy heavy ions breach spacecraft shielding materials, they fracture into a slew of lesser ions with lower atomic numbers that can go deeper than the incoming particles. These nuclei in GCR can have enough energy to pierce any shielding device currently employed on mission spacecraft. Interactions among the ions in the space environment are vital in various areas critical for space study and exploration due to the secondary radiation fields they generate. Space radiation may infiltrate housing, ships, equipment, and spacesuits, thereby putting personnel at risk. One of the most difficult issues in keeping astronauts fit and healthy while travelling through the solar system is minimising the physiological changes produced by space radiation exposure~\cite{calc06}. Due to the difficulty of reproducing GCR radiation environments on Earth, stochastic 
({Monte Carlo}) or deterministic transport algorithms are required for risk assessment of exploratory mission scenarios and shielding design. Radiation transport algorithms employ fragmentation cross sections, also known as production cross sections, which include charge-changing cross sections, neutron removal cross sections, interaction cross sections, and total reaction cross sections to measure this phenomenon. The primary aim is to accurately model the interaction between particles and materials in space environments. The nuclear fragmentation cross sections used in space radiation transport codes at NASA are derived from experimental data and theoretical models. Experimental measurements~\cite{zet07,zet08,zet11,not07,iancu05} of nuclear fragmentation cross sections involve bombarding a target material with a beam of energetic particles and analysing the resulting nuclear reactions. These experiments are typically performed using accelerators and detectors to measure the products of the interactions. Experimental data provide crucial information on the probabilities of various nuclear reactions and fragment production for different projectile energies and target materials. On the theoretical side, the required cross sections are computed using some semi-empirical models like the Silberberg-Tsao~\cite{tsao} model, and a series of codes like NUCFRAG~\cite{nuc1}, NUCFRG2~\cite{nuc2}, NUCFRG3~\cite{nuc3} developed by NASA. There are also a few software packages that are used in cross section estimation for space radiation applications like GEANT4~\cite{geant}, FLUKA~\cite{fluka}, PHITS~\cite{phits} and LISE++~\cite{lise}.

A proper analysis of the fragmentation process and cross section measurements are also useful ingredients for modeling the transport code for the use of ion beam therapy dose calculations and treatment planning. When charged particles travel through matter, they lose energy largely through interactions with the material's atomic electrons and nucleus. The depth of penetration of the particles affects the energy deposition. The Bragg curve describes the connection between particle energy deposition or stopping power and material penetration depth~\cite{brag19,prot15}. When the Bragg peak is considered, the interaction cross section or total reaction cross section directly affects the total energy deposition profile of the charged particles~\cite{luhr2012impact,luoni2021total}. A higher value of cross sections provides a wider and shallower Bragg peak. This is due to the fact that particles with bigger cross sections have more frequent contact, resulting in a more progressive loss of energy across a wider depth range. The values of the cross sections vary from ion to ion depending on the nuclear interactions the ion undergoes when travelling through the material or human body. The fragments or ions like $^4$He and $^{12}$C become very useful for radiation therapy because of their favourable depth to dose Bragg curve. Recent results~\cite{he3,ox12} on $^3$He and $^{12}$O ions have also proven to be very useful in heavy ion beam therapy.

Nuclear fragmentation in heavy-ion collisions at intermediate energies has been a subject of extensive research over the past few decades~\cite{tanihata2013recent,chatterjee2018breakup,bonaccorso2018direct}. 
Of particular interests are exotic nuclei, such as halo nuclei~\cite{tanihata2013recent,chatterjee2003,chatterjee2000projectile} with extended neutron distributions or bubble nuclei~\cite{Vishal2020,Grassobubble2009} with reduced central densities, which exhibit unique structural properties that influence reaction dynamics. These exotic features not only provide valuable insights into nuclear interactions but also play a critical role in understanding nucleosynthesis and the evolution of matter in astrophysical environments~\cite{barman2024effect,chatterjee2020status}, such as supernovae and neutron star mergers.

The charge-changing cross sections across nuclei can be effectively described using the abrasion-ablation framework~\cite{Huf75,Carl92,Scheidenberger,tanaka2022}. This process involves two principal mechanisms where proton removal plays a central role. In one scenario, abrasion of protons occurs during the initial impact, often accompanied by neutron removal, followed by neutron evaporation in the later stage. In the alternative scenario, neutrons are primarily removed during the initial phase, while subsequent evaporation leads to the loss of one or more protons along with additional neutrons. These distinct pathways collectively influence the final charge of the residual nucleus.
In the abrasion step, nucleons from the overlapping zone of the projectile and target are abraded, depending on the impact parameter, resulting in the formation of excited remnants of the initial nuclei, known as prefragments. In the subsequent ablation step, these prefragments deexcite through the evaporation of light particles. The evaporation chain continues until the excitation energy of the prefragments falls below the particle emission threshold energies, resulting in the formation of the final fragments. 

In this paper, we study charge-changing cross sections using the abrasion-ablation model for $^{12}\text{C}$, $^{14}\text{N}$, $^{16}\text{O}$, and $^{20}\text{Ne}$ projectiles on elemental targets C, Al, and Cu at a beam energy around 290 MeV/nucleon. The abrasion stage, representing direct nucleon removal, is modelled using the Glauber formalism~\cite{glauber1959,Bert21}  with appropriate nuclear density distributions. The ablation stage is governed by statistical evaporation models~\cite{lise} based on Hauser-Feshbach theory, which calculates the light particle evaporation probability from excited prefragments. 

To assess the evaporation contributions, we compare statistical model calculations with estimates from experimental charge-changing cross section data by subtracting the calculated abrasion component. The evaporation process depends on several key factors, including excitation energy distributions (EED), nuclear level densities, decay width ratios, and neutron/proton separation energies of the prefragment nuclei. A detailed analysis is performed to examine the influence of EED parameters across different targets, revealing a linear inverse correlation between the maximum excitation energy and the projectile neutron separation energy.

The paper is organized as follows. The theoretical formalism involving the abrasion-ablation formalism is briefly described in \ref{sec:level2}, followed by the details of the results and discussions in \ref{sec:level3}, and the summary and conclusions of the present work in \ref{sec:level3}.
\section{\label{sec:level2}Theoretical Formalism}
\subsection{Primary fragmentation or abrasion stage}
The Glauber multiple scattering approach~\cite{glauber1959} is a credible theoretical model for calculating projectile fragmentation in high-energy collisions. Let a projectile of mass number $A_P$ that consists of $N_P$ neutrons and $Z_P$ protons produce a fragment of mass number $A_F$ with neutron number $N_F$ and proton number $Z_F$ in the collision with a target nucleus of mass number $A_T$. The differential cross-section for the primary yield of the fragment with $N_F$ neutrons and $Z_F$
protons are written as the product of the density of states $\omega(E, Z_F, A_F)$ and an integral over impact parameter $b$~\cite{Huf75,Carl92,Bert21},
\begin{widetext}
\begin{equation}
\frac{d\sigma}{dE}(E,Z_F,A_F) = \omega(E,Z_F,A_F) \int d^2b \, \left[P_p(\mathbf{b})\right]^{Z_F} \left[P_n(\mathbf{b})\right]^{N_F}  \left[1 - P_n(\mathbf{b})\right]^{N_P - N_F}  \left[1 - P_p(\mathbf{b})\right]^{Z_P - Z_F}.
\label{eq:diff_cross_section}
\end{equation}
\end{widetext}

The integral represents the cross section for each primary fragment state as a sum over impact parameters, considering the probability that \( Z_F \) protons and \( N_F = A_F - Z_F \) neutrons from the projectile do not scatter, while the remaining ones do. The terms \( P_p(\mathbf{b}) \) and \( P_n(\mathbf{b}) \) denote the probabilities that a proton or neutron from the projectile, respectively, does not collide with the target. $\omega(E,Z_F,A_F)$ is the density of states, which is calculated by counting all combinations of projectile holes consistent with the fragment's charge and neutron numbers~\cite{Huf75}. Thus, the energy-integrated primary cross section \( \sigma(E, Z_F, A_F) \) inherently includes the bionomial factor \( \binom{Z_P}{Z_F} \binom{N_P}{N_F} \), a component that has been incorporated in several previous models~\cite{Huf75, Oliveira1979}.
\begin{equation}
    \sigma(E,Z_F,A_F) = \binom{N_P}{N_F} \binom{Z_P}{Z_F} \int d^2b \, 
    \left[P_p(\mathbf{b}) \right]^{Z_F} \left[ P_n(\mathbf{b}) \right]^{N_F}  
    \left[ 1 - P_n(\mathbf{b}) \right]^{N_P - N_F}  
    \left[ 1 - P_p(\mathbf{b}) \right]^{Z_P - Z_F}.
\label{eq:frag_cross_section}
\end{equation}
Now following the work in Ref.~\cite{teixeira2022nuclear}
we can calculate the required cross sections as follows.

The total reaction cross section is obtained by taking the sum of cross sections in all events where at least one nucleon is removed, which is given as 
\begin{eqnarray}
 \sigma_R= \int d^2b \left[1-[P_p(\textbf{b})]^{Z_P}[P_n(\textbf{b})]^{N_P}\right].
\end{eqnarray}
The partial neutron removal cross section ($\sigma_{-xn}$) can be calculated from Eq. (\ref{eq:frag_cross_section}) by setting $Z_F=Z_P$. The cross section for removal of $x$ neutrons from the projectile 
\begin{equation}
\sigma_{-xn} = \binom{N_P}{x} \int d^2b \, [P_p(\mathbf{b})]^{Z_P} [P_n(\mathbf{b})]^{N_P - x} [1 - P_n(\mathbf{b})]^x.
\label{eq:neutron_removal}
\end{equation}
The total neutron removal cross section is calculated by adding all events with at least one neutron removed without disturbing the proton number in the projectile
\begin{eqnarray}
\sigma_{\Delta N}&=&\sum_{N_F=0}^{N_P-1} \sigma(N_F,Z_P)\nonumber\\
&=& \int d^2b [P_{p}(\textbf{b})]^{Z_P} \left\{1-[P_n(\textbf{b})]^{N_P}\right\}.
\end{eqnarray}
The charge-changing cross section in the abrasion is measured by taking all the events in which at least one proton is removed and is written by subtracting the total neutron removal cross section from the total reaction cross section,
\begin{eqnarray}   \sigma_{\rm cc}&=&\sigma_R-\sigma_{\Delta N}\nonumber\\
    &=& \int d^2b \left\{1-[P_p(\textbf{b})]^{Z_P}\right\}.
    \label{Eq:charge_cc}
\end{eqnarray}
The probability $P_p(\textbf{b})$ can be determined from the S-matrix concept using the optical potential. The eikonal S-matrix is defined as $\mathcal{S}_p(\textbf{b})=\exp[i\chi_p(\textbf{b})]$, for a projectile proton scattering from a target nucleus, where $\chi_p(\textbf{b})$ is the eikonal phase shift function \cite{Bert21}.
Then the eikonal phase shift function with the optical limit approximation~\cite{Terashima2014,Suzuki2016}
\begin{equation}
\chi_p(\mathbf{b}) = i \int d^2s \, \rho_{pz}^P(\mathbf{s}) \int d^2t \, \Big[ Z_T \rho_{pz}^T(\mathbf{t}) \Gamma_{pp}(\mathbf{b} + \mathbf{s} - \mathbf{t}) + N_T \rho_{nz}^T(\mathbf{t}) \Gamma_{pn}(\mathbf{b} + \mathbf{s} - \mathbf{t}) \Big].
\end{equation}
with the thickness function $\rho_{Nz}^{P(T)}$ related to the density by, {\it e.g.},
\begin{equation}
    \rho_{Nz}^{P(T)}=\int \rho_N^{P(T)}(\sqrt{\textbf{s}^2+z^2})dz,  ~~N=p,n
\end{equation}
where $z$ is the coordinate along the direction of the incident beam.
The densities of the projectile(target) - proton(neutron) are 
$\rho_{p(n)}^{P(T)}(r)$ and are normalised so that $\int \rho_{p(n)}^{P(T)}(r) d^3{r}=1$.
Then the S-matrix
\begin{equation}
\mathcal{S}_p(\mathbf{b}) = \exp\left(- \int d^2s \rho_{pz}^P( \mathbf{s}) \int d^2t \left[ Z_T \rho_{pz}^T(\mathbf{t}) \Gamma_{pp}(\mathbf{b} + \mathbf{s} - \mathbf{t}) + N_T \rho_{nz}^T(\mathbf{t}) \Gamma_{pn}(\mathbf{b} + \mathbf{s} - \mathbf{t}) \right] \right).
\label{Eq:phase_shift}
\end{equation}
$\textbf{s}~ (\textbf{t})$ represents the two-dimensional coordinate of a particular projectile (target) nucleon relative to the center of mass of the projectile (target) nucleus, which lies on the plane perpendicular to the incident momentum of the projectile.
The profile function $\Gamma_{pN}(\textbf{b})$ $(N=p,n)$ describes the scattering of projectile protons (neutrons) with the target protons (neutrons), which is often parametrized in a finite range~\cite{abu2008reaction} as
\begin{equation}
\Gamma_{pN}(\mathbf{b}) = \frac{1 - i\alpha_{pN}}{4\pi\beta_{pN}} \, \sigma_{pN}^{ \rm tot} \, \exp\left(-\frac{\mathbf{b}^2}{2\beta_{pN}}\right).
\label{Eq:profile_function}
\end{equation}
The parameters used in the proton-nucleon $(pN)$ scattering model are defined as follows: 
\(\alpha_{pN}\) is the ratio of the real to the imaginary part of the $pN$ scattering amplitude in the forward direction,\(\sigma_{pN}^\text{tot}\) denotes the total $pN$ cross section, \(\beta_{pN}\) is the slope parameter of the $pN$ differential cross-section, characterizing the spatial distribution of the scattering amplitude. The proton-proton (Coulomb removed) and proton-neutron total cross sections ($\sigma_{pN}^{ \rm tot}$)  parameters are generally obtained from a fit of experimental data. The standard parameter set for proton-neutron and proton-proton scattering can be found in tabulated form in Ref.~\cite{abu2008reaction}. The probability in Eq. (\ref{eq:diff_cross_section}) is then calculated $P_p(\textbf{b})=\left|\mathcal{S}_p(\textbf{b})\right|^2$. Similarly, the probability $P_n(\textbf{b})=\left|\mathcal{S}_n(\textbf{b})\right|^2$, 
given by interchanging the role of a proton with a neutron$(p\leftrightarrow n)$ in Eq.~(\ref{Eq:phase_shift}) and Eq.~(\ref{Eq:profile_function}).

\subsection{Secondary fragmentation or ablation stage}

In the ablation stage~\cite{Oliveira1979,Huf75,Carl92,GS90,Scheidenberger}, the final fragmentation products are formed by deexciting the residual fragment in the abrasion stage through the
evaporation of light particles. Here, we are interested in the process where neutrons are removed during the initial abrasion stage, followed by the evaporation of protons during the final stage. 

The proton particle evaporation contribution is calculated following the development by   Refs.~\cite{Scheidenberger,tanaka2022}
 \begin{equation} \sigma_{\rm cc}^{ {\rm evap}}=\sum_{x=1}^{N_P} \sigma_{-xn}P_{-xn},
 \label{Eq:sig_evap}
 \end{equation}
 with $\sigma_{-xn}$ as the partial neutron removal cross section in the abrasion stage as formulated in Eq. (\ref{eq:neutron_removal}). $P_{-xn}$ as the corresponding charged particle evaporation probability after initial neutron removal, that depends on the excitation energy distribution (EED) of the prefragment and the probability to evaporate at least one proton at a certain excitation energy as,
 \begin{equation}
     P_{-xn}=  \int_0^\infty w_{-xn}(E_{\text{ex}}) f(E_{\text{ex}}, A_P - x, Z_P) \, dE_{\text{ex}}
     \label{Eq:prob_p_in}
 \end{equation}
$E_\mathrm{ex}$ is the excitation energy of the residual fragment after $x$ neutron removal, $f(E_{ \rm ex},A_p-x, Z_p)$ is the proton evaporation probability (same as the branching ratio) at certain excitation energy, $w_{-xn}(E_ {\rm ex})$ is the excitation energy distribution (EED) of the prefragment. The  Gaimard-Schmidt(GS) approach~\cite{GS90} is used to calculate the EED of the residual fragment as in Ref.~\cite{tanaka2022,Scheidenberger}.
The LISE++ (GEMINI++)~\cite{Charity, Mancusi2010} code calculates the charge particle evaporation probability ($f(E_{ \rm ex},A_P- x,Z_P)$)  at certain excitation energy based on the Hauser-Feshbach theory and Weisskopf-Ewing formalism.
\section{\label{sec:level3}Results and discussions}
In this section, we briefly present the results of charge-changing cross sections (Eq.~(\ref{Eq:charge_cc})) calculated in the abrasion stage that uses the Glauber model with realistic density inputs. We then calculate proton evaporation contributions in the ablation stage and analyze their correlation with key influencing factors related to the evaporation process.

\subsection{Single-nucleon density inputs}
\begin{figure}
    \centering  \includegraphics[width=0.6\columnwidth]{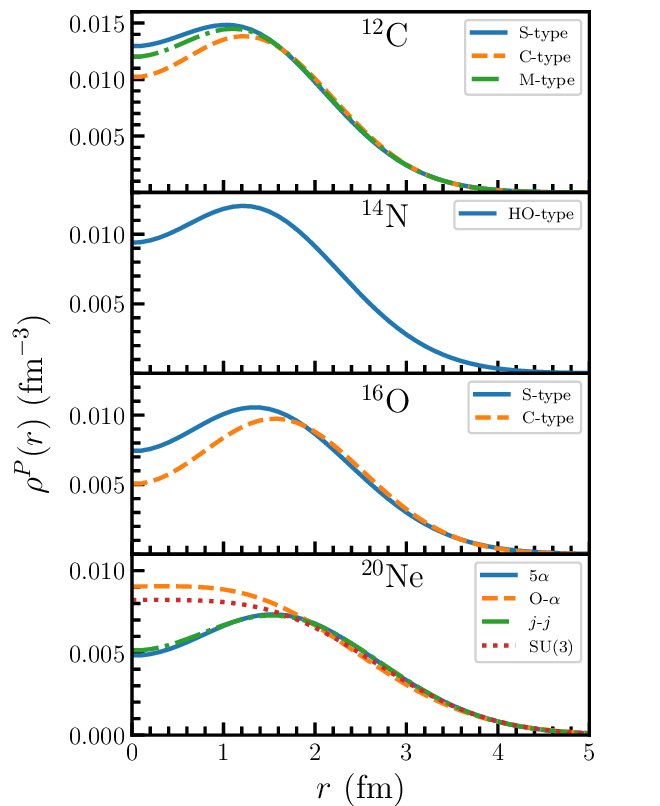}
    \caption{Single-nucleon density distributions for \(^{12}\text{C}\), \(^{14}\text{N}\), \(^{16}\text{O}\), and \(^{20}\text{Ne}\), 
    presented from top to bottom, respectively. Different nuclear density configurations 
     taken from Ref.~\cite{Horiuchi2023} for $^{12}$C and $^{16}$O and Ref.~\cite{yamaguchi2023} for $^{20}$Ne  are
considered, as discussed in the text.}
\label{fig:density}
\end{figure}
We now show a few input density configurations for $^{12}$C, $^{14}$N, $^{16}$O, and $^{20}$Ne in Fig.~\ref{fig:density}, considered in the literature. For \(^{12}\text{C}\) and \(^{16}\text{O}\), the single-nucleon densities are calculated based on the 3\(\alpha\) and 4\(\alpha\) cluster configurations (C-type), respectively. A shell model type (S-type) density is also considered for both nuclei. It is also well known that \(^{12}\text{C}\) represents a mixture of shell and cluster configurations \cite{otsuka2022,Chernykh2007}. To account for this mixture, the mixed configuration (M-type) is additionally considered for \(^{12}\text{C}\). The shell and cluster density profiles are derived by the approach outlined in \cite{Horiuchi2023,Horiuchi2022} and tailoring it for the current multi-\(\alpha\) scenarios. Details of the theoretical framework underlying this calculation are provided in \cite{Horiuchi2022}. The S-type and C-type configurations show significantly different density profiles, despite having the same root-mean-square radii. The density profile near the surface of the C-type configuration declines more steeply than that of the S-type configuration, which may be more effective for peripheral nucleon removal. 

The density of $^{14}$N is calculated using shell model configuration with a harmonic oscillator (HO) basis. 

For \(^{20}\text{Ne}\), two cluster configurations (5\(\alpha\) and \(^{16}\text{O}+\alpha\)) and two shell-model configurations ($j$-$j$ coupling and SU(3)) are examined. 
The $j$-$j$ coupling configuration corresponds to $(0d_{5/2})^4$, where four nucleons occupy the $0d_{5/2}$ orbital, while the SU(3) configuration corresponds to $(1s0d)^4$, involving nucleons in the $1s$ and $0d$ shells based on SU(3) symmetry. 

These different density configurations have been systematically studied through calculations of angular distributions in proton elastic scattering, as reported in Refs.~\cite{Horiuchi2023,Horiuchi2022,yamaguchi2023}. Based on these studies, the most appropriate density configurations are the Mixed type (M-type) for $^{12}$C, a 4$\alpha$ cluster type (C-type) for $^{16}$O, and a $^{16}$O + $\alpha$ (O-$\alpha$ type) bicluster structure for $^{20}$Ne. Our study will use these specific density distributions for cross section calculations in the abrasion step, ensuring consistency with the most appropriate structure models identified from proton elastic scattering studies.

\subsection{ Direct fragmentation or abrasion stage}

\begin{table}
\centering
\caption{\label{tab:table1}Charge-changing cross sections in the abrasion stage of projectile fragmentation for 
$^{12}$C, $^{14}$N, $^{16}$O, and $^{20}$Ne on elemental targets C, Al, and Cu at a beam 
energy of 290 MeV/nucleon. Calculations were performed using the finite-range optical limit 
approximation (FROLA) with the most appropriate nuclear density inputs: mixed-type (M-type) 
density for $^{12}$C, a shell model with a harmonic oscillator basis for $^{14}$N, 
C-type (4$\alpha $ cluster) density for $^{16}$O, and a $^{16}$O + $\alpha$ cluster configuration for $^{20}$Ne. All cross sections in this table are rounded off to the nearest integer.}

\setlength{\tabcolsep}{5pt} 
\renewcommand{\arraystretch}{0.8} 
\begin{tabular}{cccc}
\hline
\hline
{Projectile} & {Target} & {Projectile} & $\sigma_{\rm cc}$ (mb) \\ 
{Nucleus}    & {Nucleus} & {Density Type} & {(Abrasion)} \\ \hline
$^{12}$C & C & {M-type} & 650\\   
& Al &  & 1001\\  
& Cu &  & 1568\\
$^{14}$N & C & {HO-type} & 720\\   
& Al &  &1070\\  
& Cu &  &1648\\ 
$^{16}$O & C & {C-type} & 772\\  
& Al &  &1130\\ 
& Cu &  & 1722\\ 
$^{20}$Ne & C & O-$\alpha$ type& 911\\  
& Al &  &1308\\ 
 & Cu &  & 1951\\ 
\hline
\hline
\end{tabular}
\end{table}

\begin{figure*} 
\centering  \includegraphics[width=\textwidth]{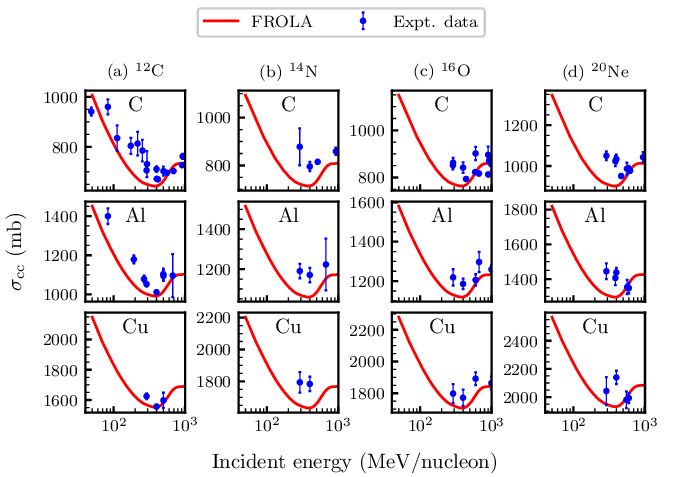}
\caption{Charge-changing cross section as a function of incident energy of the projectile nucleus for \(^{12}\text{C}\), \(^{14}\text{N}\), \(^{16}\text{O}\), and \(^{20}\text{Ne}\) on elemental targets: Carbon (C), Aluminium (Al), and Copper (Cu). The plots are organized into four columns corresponding to different projectiles: (a) \(^{12}\text{C}\), (b) \(^{14}\text{N}\), (c) \(^{16}\text{O}\), and (d) \(^{20}\text{Ne}\). Each column presents the results for the respective projectiles on three targets: Carbon (C), Aluminium (Al), and Copper (Cu). The solid line represents theoretical calculations using the FROLA, and experimental data for the charge-changing cross sections in the energy range from 40 to 1000 MeV/nucleon are included, which are taken from Refs. \cite{zet11, Alpat2013, Tran2016, schall1996charge, ferrando1988measurement, chulkov2000total, Golovchenko2002, cecchini2008fragmentation, webber1990} for \(^{12}\text{C}\), from Refs. \cite{zet11, webber1990, chulkov2000total, schall1996charge} for \(^{14}\text{N}\), from Refs. \cite{yamaguchi2011, webber1990, ferrando1988measurement, chulkov2000total, zet11, schall1996charge} for \(^{16}\text{O}\), and from Refs.~\cite{webber1990, zet11, cheng2012, zhang2012} for \(^{20}\text{Ne}\).}
\label{fig:12C_elements}
\end{figure*}

Table~\ref{tab:table1} presents the charge-changing cross section for projectile nuclei $^{12}\mathrm{C}$,$^{14}$N,$^{16}$O, and $^{20}\mathrm{Ne}$ on elemental targets such as carbon (C), aluminum (Al), and copper (Cu) with beam energy 290 MeV/nucleon, during the direct fragmentation (abrasion) stage. The cross sections calculated using the finite-range optical limit approximation (FROLA) correspond to the most appropriate density configurations, as discussed in the previous section. These cross sections show slight variations when compared to other density configurations, with differences generally within $2\%$.

Fig.~\ref{fig:12C_elements} shows the FROLA calculation for the charge changing cross sections ($\sigma_{\rm cc}$) on the elemental targets, as a function of beam energy. 
The results in the abrasion stage ($\sigma_{\rm cc}$) systematically underestimate charge-changing cross sections compared to experimental data, as also observed in previous studies~\cite{yamaguchi2011,tanaka2022}. Calculations of charge-changing cross sections using the Glauber model indicate that not only the protons in the projectile but also the presence of neutrons play a significant role in explaining the observed data. 

Some studies~\cite{yamaguchi2011,yamaguch2010,Gambhir2004} propose using a phenomenological energy dependent or $Z/N$ ratio dependent correction parameter that accounts for the contribution of neutrons in the projectile, allowing for a more precise description of the data. On the other hand, the Glauber model only considers the direct removal of protons from the projectile, neglecting secondary processes. In reality, charged particles, predominantly protons, can be emitted following the direct removal of neutrons, especially when the prefragment is in a highly excited state~\cite{Scheidenberger,tanaka2022,zhao2023}. When neutrons are removed from the projectile during the abrasion stage (direct fragmentation), the residual nucleus may become excited to states beyond the particle evaporation threshold. In such cases, deexcitation occurs through the removal of charged particles, such as protons or alpha particles. This process contributes to the overall charge changing cross section, increasing the likelihood of charge changing reactions beyond the initial neutron removal.  
This motivates us to explore the proton evaporation contribution to the charge changing process on various elemental targets (C, Al, Cu) in the next section.

\subsection{Proton evaporation contributions}

Estimating proton evaporation contributions in the secondary fragmentation or ablation stage is done in two ways. 
The first involves evaporation contributions calculated following Eq.~(\ref{Eq:sig_evap}) and Eq.~(\ref{Eq:prob_p_in}). The second is subtracting the direct process (estimated with the FROLA) from the experimental data. A comparison of these two methods would then be made at the end. 

Eq.~(\ref{Eq:sig_evap}) involves calculations of partial neutron removal cross section ($\sigma_{-xn}$) and  proton evaporation probability ($P_{-xn}$). The probability, in turn, depends on key factors like excitation energy distributions (EED) of the prefragment and probability at a particular excitation energy. We now proceed to present the calculations of these factors in detail.

\subsubsection{Excitation energy distribution (EED)}

\begin{figure} 
    \centering  \includegraphics[width=0.7\columnwidth]{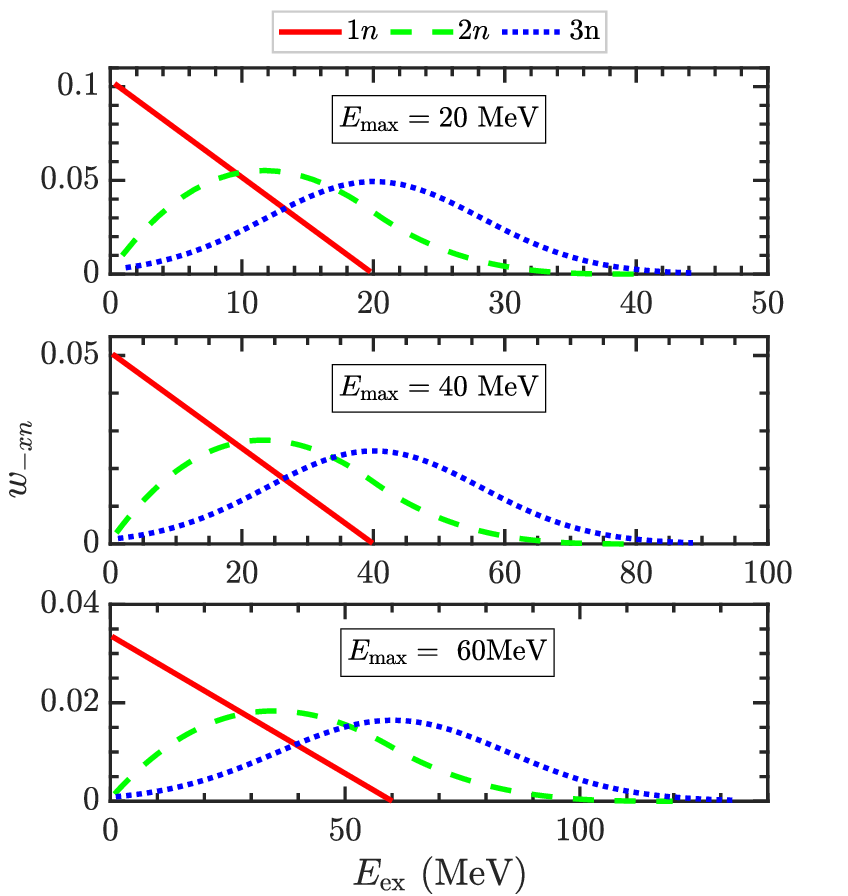}   \caption{Excitation energy distributions of the prefragment after the removal of \( x = 1 \)-3 neutrons in direct fragmentation(abrasion stage), shown for \( E_{\text{max}} = 20 \), 40, and 60 MeV, respectively.}
   \label{fig:w_xn}
\end{figure}

The excitation energy distribution (EED) \( w_{-xn}(E_{ex})\) of the prefragment, which is an important parameter in Eq.~(\ref{Eq:prob_p_in}) is calculated using the LISE++ simulation code~\cite{lise}. LISE++ offers several theoretical approaches for determining the EED, with the Gaimard-Schmidt (GS)~\cite{GS90} model being one of the primary options. The prefragment excitation energy is determined by summing the energies of the hole states remaining in the core of the initial nucleus after nucleon removal. GS model adopts a simple approach, that uses a  functional shape corresponds to an approximation of the single hole state density in the Woods-Saxon potential, as done in Refs.~\cite{Scheidenberger,tanaka2022} requiring only two input parameters: the number of nucleons removed and the maximum excitation energy ($E_\text{max}$) of the prefragment formed following a one-nucleon removal reaction. $E_\text{max}$ plays a crucial role in determining the excitation energy of prefragments in nuclear reactions. It represents the maximum energy that can be deposited by a single removed nucleon into the prefragment. Fig.~\ref{fig:w_xn} illustrates the excitation energy distribution of the prefragment following the removal of one, two, and three nucleons from the nucleus. The distributions are characterized by the maximum excitation energy, which is set at 20 MeV, 40 MeV, and 60 MeV for the respective cases, for example. For one neutron abrasion case the distribution is normalized and decreases linearly with increasing excitation energy until \( E_{\text{max}} \), where it abruptly drops to zero. This behavior reflects the simplified assumption that the probability of excitation decreases as energy approaches the maximum limit. As more neutrons are removed, the excitation energy distribution generally becomes 
broader, more complex, and
shifted towards higher energies. The mean energy~\cite{Scheidenberger} per removed nucleon  $\langle E \rangle =  ({x}/{x + 1}) E_{\text{max}}$, which gives 10 MeV for first neutron removal for $E_\text{max}$= 20 MeV and consequently this value increases and become maximum for higher neutron removal.

\subsubsection{Proton evaporation probability}
\begin{figure}
\centering  \includegraphics[width=0.6\columnwidth]{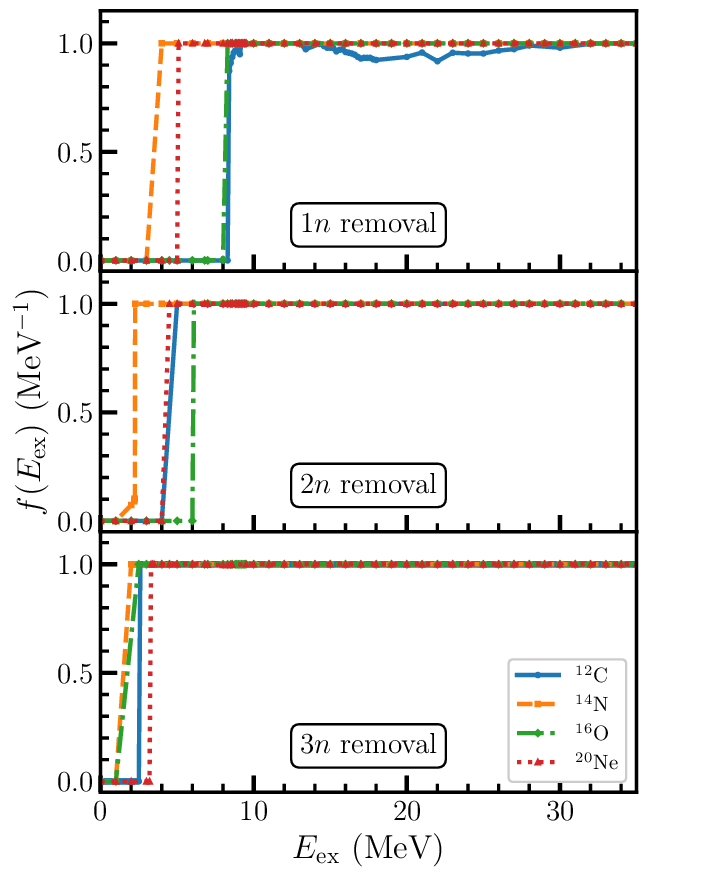}
\caption{Probability $(f(E_\text{ex},A_P-x,Z_P))$ of proton evaporation after initial removal of neutrons ($x=1$-3) from the projectiles $^{12}$C, $^{14}$N, $^{16}$O, and $^{20}$Ne as a function of prefragment excitation energy ($E_\mathrm{ex}$).}
   \label{fig:fx_Eex}
\end{figure}

\begin{figure}
    \centering  \includegraphics[width=0.7\columnwidth]{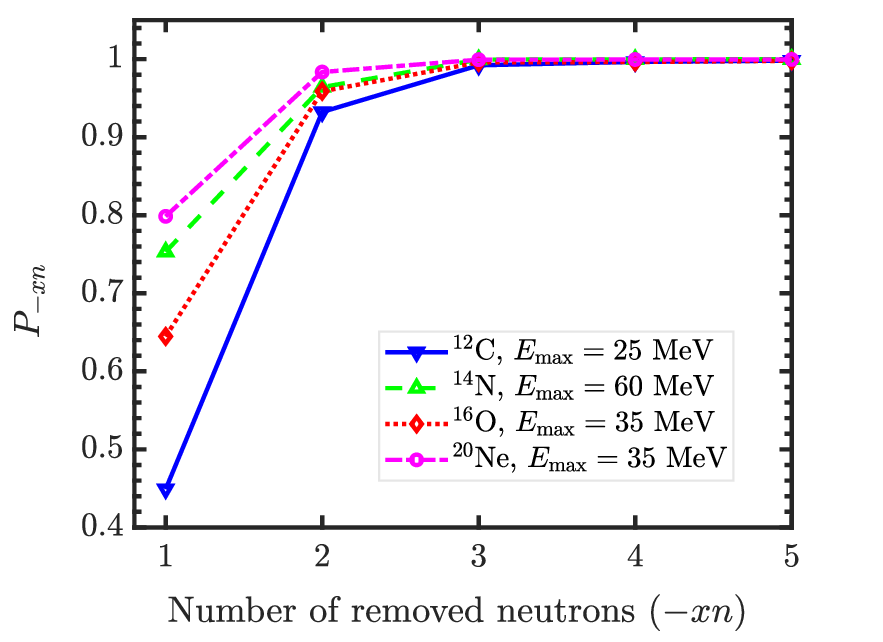}   \caption{Proton evaporation probability (\( P_{-xn} \)) as a function of the number of neutrons removed in the initial fragmentation (abrasion stage). The maximum excitation energy parameter \( E_\mathrm{max} \) is selected to approximately reproduce the charge-changing cross-section data at 290 MeV/nucleon for a carbon target.
}
   \label{fig:p_xn}
\end{figure}

We now describe the calculation of the probability of proton evaporation ($f(E_\text{ex},A_P-x,Z_P)$) as a function of excitation energy. The calculations are performed using the LISE++ (GEMINI++) simulation code~\cite{lise,Charity,Mancusi2010}, which incorporates the Hauser-Feshbach and Weisskopf formalisms for compound nucleus decay. This approach accounts for decays involving the emission of at least one proton during cascade deexitation. Fig.~\ref{fig:fx_Eex} illustrates the variation in the charged particle evaporation probability as a function of excitation energy (\(E_{\text{ex}}\)) for the prefragments of \(^{12}\text{C}\), \(^{14}\text{N}\), \(^{16}\text{O}\), and \(^{20}\text{Ne}\) following the removal of one, two, and three neutrons during the abrasion stage. The evaporation probability curve begins to rise at or near the energy that equals to proton separation energy plus the Coulomb barrier for the prefragments. This is the excitation energy threshold beyond which the prefragment is sufficiently excited for charged particle evaporation processes to occur. After this energy threshold, the prefragment nucleus loses at least one proton, which contributes to the charge changing cross section. This trend is also followed by the other prefragments produced after the removal of two or three neutrons in the abrasion stage.
With these values of $f(E_\text{ex})$ and ${w}_{-xn}(E_\text{ex})$, the evaporation probability ($P_{-xn}$) is calculated and shown in Fig.~\ref{fig:p_xn} for up to 5 neutron removals from the projectile. The parameter $E_\mathrm{max}$ is fixed to a value that reproduces the estimated evaporation cross sections from the experimental data at beam energy 290 MeV/nucleon on carbon target. The variation of \( P_{-xn} \) indicates that as the number of neutrons removed from the projectile increases, the charged particle evaporation probability also increases. For all projectile nuclei, one notices this probability tends to 1 when the $Z/N$ ratio becomes greater than 1, reflecting a direct relationship between the degree of neutron removal and the likelihood of charged particle evaporation, contributing significantly to the charge-changing cross section. 

\begin{figure}
    \centering  \includegraphics[width=0.7\columnwidth]{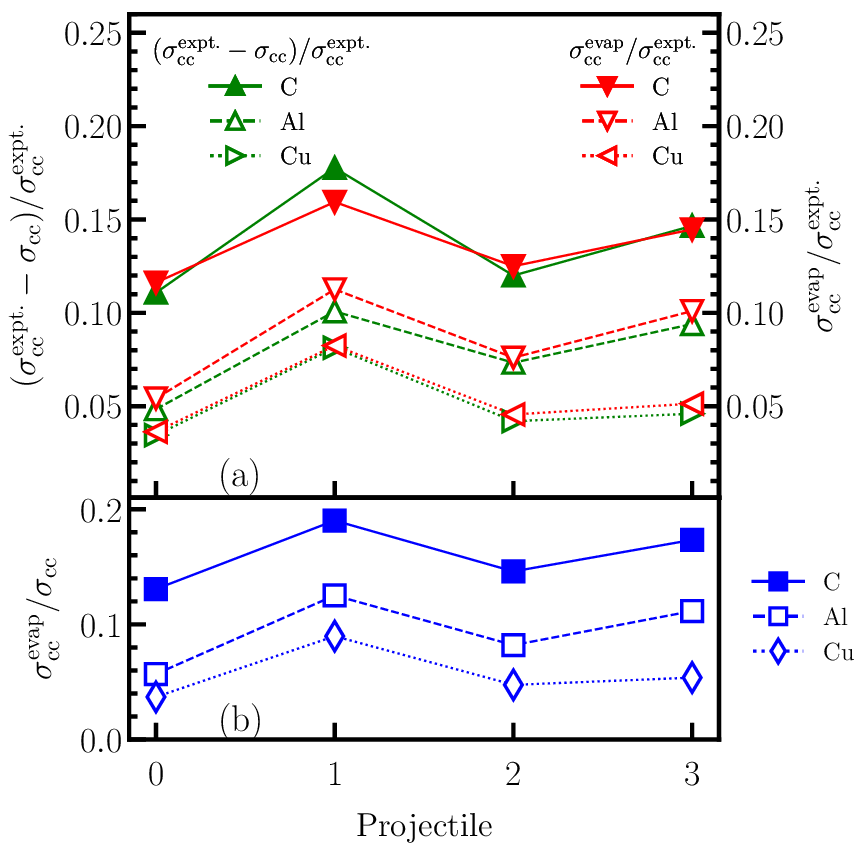}  \caption{Top panel ({a}) represents the ratio of the estimated evaporation contributions $(\sigma_\mathrm{cc}^{\text{expt.}} - \sigma_\mathrm{cc})$ to the total experimental charge-changing cross section $(\sigma_{\rm cc}^{\text{expt.}})$ on the left vertical axis, while the ratio $\sigma_\mathrm{cc}^{\text{evap}} / \sigma_{\rm cc}^{\text{expt.}}$ is shown on the right vertical axis. The bottom panel ({b}) displays the ratio $\sigma_\mathrm{cc}^{\text{evap}} / \sigma_\mathrm{cc}$. Distinct marker styles differentiate the targets, with filled markers representing carbon and unfilled markers denoting aluminum and copper. Individual data points are marked, while lines serve as visual guides to indicate trends across different projectiles and targets.}
  \label{fig:sigma_ratio}
\end{figure} 

\begin{figure}
    \centering  \includegraphics[width=0.7\columnwidth]{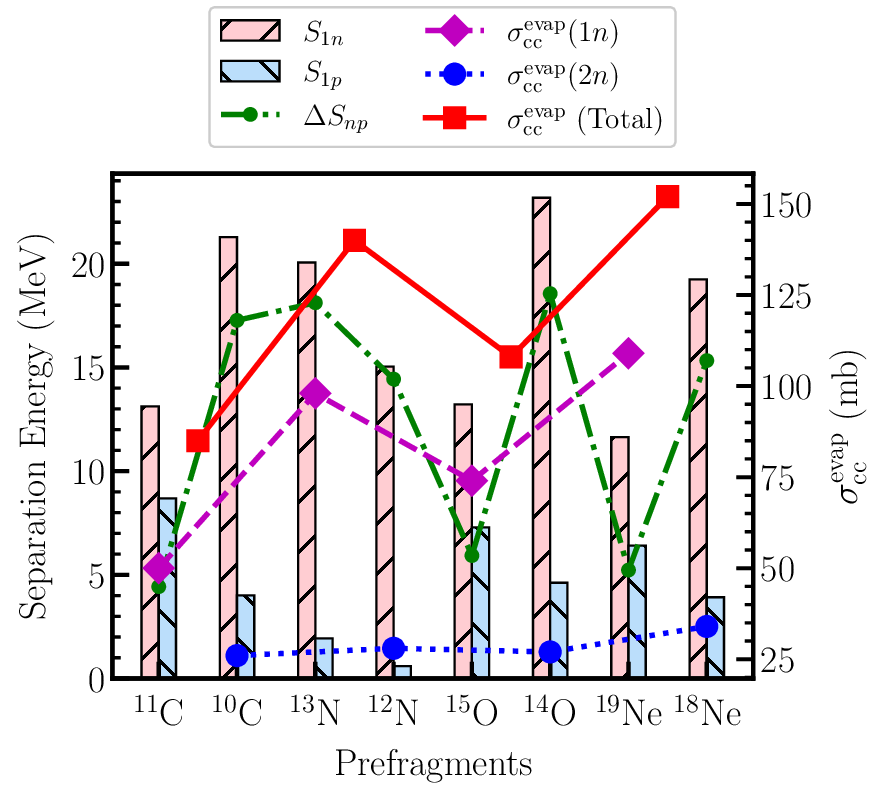}\caption{Separation energies and evaporation cross sections for various prefragment nuclei. The neutron separation energy (\( S_{1n} \)) is represented by bars with forward-slash patterned shading, while the proton separation energy (\( S_{1p} \)) is depicted with backslash-patterned shading. The difference between neutron and proton separation energies (\( \Delta S_{np} \)) is shown as a dashed-dotted line. Evaporation cross sections following one-neutron removal in the abrasion stage (\( \sigma_\mathrm{cc}^\mathrm{evap}(1n) \)) are represented by diamond markers with dashed lines, two-neutron removal (\( \sigma_\mathrm{cc}^\mathrm{evap}(2n) \)) by circular markers with dotted lines, and total evaporation cross sections (\( \sigma_\mathrm{cc}^\mathrm{evap}(\text{Total}) \)) by square markers with solid lines. The separation energy data are calculated using the atomic mass table from Ref. \cite{ame2020}.}  \label{fig:proton_evap}
\end{figure} 

\begin{figure}
    \centering  \includegraphics[width=0.7\columnwidth]{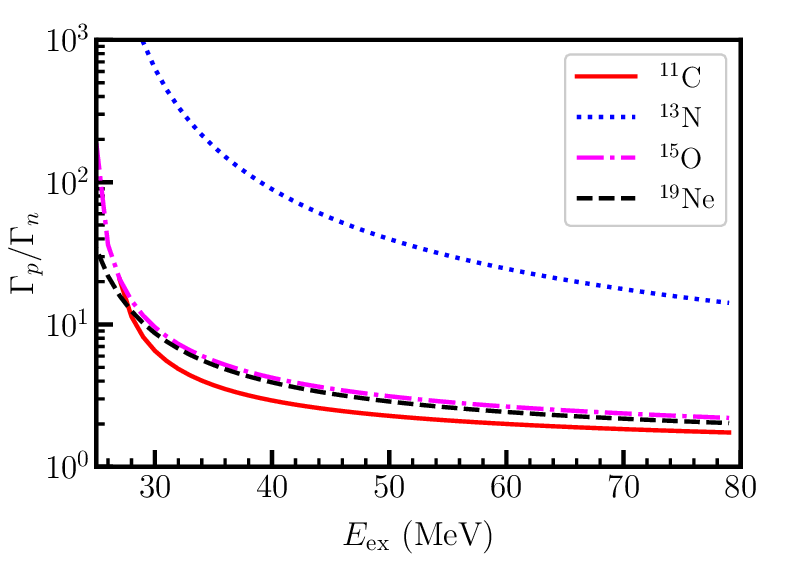}  \caption{ Ratio of the proton to neutron decay widths as a function of the excitation energy of the prefragment after $1n$ removal from the projectiles.}  \label{fig:gamma_ratio}
\end{figure}

\begin{table}[ht]
\centering

\caption{Proton evaporation contributions ($\sigma_{\rm cc}^{\text{evap}}$) [Eq.~(\ref{Eq:sig_evap})] to charge-changing cross section for various projectile nuclei on elemental targets (C, Al, Cu) at a beam energy around 290 MeV/nucleon. The partial neutron removal cross sections ($\sigma_{-xn}$) [Eq.~(\ref{eq:neutron_removal})] for 1$n$, 2$n$, 3$n$, and beyond 3$n$ channels are also shown, along with their respective contributions to $\sigma_{\rm cc}^{\text{evap}}$. $\sigma_{\rm cc}^{\text{evap}}$ is calculated by summing up all the contributions from all neutron removal channels. For comparison, the evaporation contribution is also estimated by subtracting the Glauber charge-changing cross section [$\sigma_{\rm cc}$ (Table~\ref{tab:table1})] from the experimental data ($\sigma_{\rm cc}^{\text{expt}}$) taken from Refs. \cite{zet11,zhao2023}. All cross sections in this table are rounded off to the nearest integer.}
\label{tab:proton_evaporation}
\renewcommand{\arraystretch}{0.8} 
\resizebox{\textwidth}{!}{ 
\begin{tabular}{c c c c c c c c c c c c c} 
\hline\hline
\multirow{2}{*}{$E_{\text{max}}$ (MeV)} & \multirow{2}{*}{Nuclei} & \multirow{2}{*}{Target} & \multicolumn{4}{c}{$\sigma_{{-xn}}$ (mb)} & \multicolumn{4}{c}{Contributions to $\sigma_\mathrm{cc}^{\text{evap}}$ (mb)} & \multirow{2}{*}{$\sigma_\mathrm{cc}^{\text{evap}}$ (mb)} & \multirow{2}{*}{$\sigma_\mathrm{cc}^{\text{expt}} - \sigma_\mathrm{cc}$ (mb)} \\
\cmidrule(lr){4-7}  
\cmidrule(lr){8-11}  
 &  &  & $\sigma_{-1n}$ & $\sigma_{-2n}$ & $\sigma_{-3n}$ & $\sigma_{-xn,~(x>3)}$ & ${-1n}$ & ${-2n}$ & ${-3n}$ & ${>-3n}$ &  &  \\
\hline
25  & $^{12}$C & C   & 114  & 28  & 8   & 1  & 50  & 26  & 8   & 1  & 85  & 81  \\
15  &          & Al  & 136  & 30  & 9   & 1  & 26  & 25  & 8   & 1  & 60  & 51  \\
10  &          & Cu  & 177  & 40  & 11  & 2  & 16  & 31  & 10  & 2  & 59  & 57  \\
\hline 
60  & $^{14}$N & C   & 115  & 29  & 8   & 3  & 100  & 29  & 8   & 3  & 140 & 156 \\
25  &          & Al  & 137  & 34  & 10  & 4  & 88  & 33  & 9   & 4  & 134 & 120 \\
15  &          & Cu  & 165  & 42  & 12  & 4  & 92  & 41  & 11  & 4  & 148 & 146 \\
\hline  
35  & $^{16}$O & C   & 115  & 29  & 9   & 3  & 69  & 27  & 9   & 3  & 108 & 106 \\
20  &          & Al  & 135  & 34  & 10  & 3  & 49  & 32  & 9   & 3  & 93  & 89  \\
15  &          & Cu  & 163  & 41  & 12  & 4  & 35  & 31  & 12  & 4  & 82  & 76  \\
\hline
35  & $^{20}$Ne & C  & 137  & 35  & 11  & 4  & 106 & 34  & 11  & 4  & 155 & 154 \\
20  &           & Al  & 159  & 40  & 12  & 6  & 92  & 36  & 12  & 6  & 146 & 137 \\
10  &           & Cu  & 186  & 47  & 14  & 7  & 48  & 36  & 14  & 7  & 105 & 94  \\
\hline
\end{tabular}
} 
\end{table}

Table~\ref{tab:proton_evaporation} presents the contributions of proton evaporation ($\sigma_\mathrm{cc}^{\text{evap}}$) [Eq.(~\ref{Eq:sig_evap})] to the charge-changing cross section for various projectile nuclei ($^{12}$C, $^{14}$N, $^{16}$O, $^{20}$Ne) on different elemental targets (C, Al, Cu) at a beam energy of 290 MeV/nucleon. The table is structured to show the neutron-removal cross sections ($\sigma_{-xn}$), which are calculated in the abrasion step using Eq.~(\ref{eq:neutron_removal}), for 1$n$, 2$n$, and 3$n$ removal channels, along with their respective contributions to proton evaporation. The individual contributions from each neutron-removal channel to $\sigma_{\rm cc}^{\text{evap}}$ [Eq.~(\ref{Eq:sig_evap})] are explicitly provided. The contributions, $\sigma_{-xn,~(x> 3)}$ from neutron removal channels beyond $-3n$, such as the removal of four or more neutrons in the abrasion stage, are calculated and summed to estimate the contributions to the `$>-3n$'
 channel. The total evaporation contribution, $\sigma_{\rm cc}^{\text{evap}}$, is then determined by summing the contributions from all neutron removal channels. 
 
The last column of Table~\ref{tab:proton_evaporation} is the evaporation estimate from experimental data~\cite{zet11,zhao2023}, and is calculated by subtracting the charge-changing cross section ($\sigma_{\rm cc}$) (Table~\ref{tab:table1}) using the Glauber model. The first column of the table represents the parameter \( E_{\text{max}} \), which is varied to achieve an approximate match of $\sigma_{\rm cc}^{\text{evap}}$ with the estimated value extracted from the experimental data.

Fig.~\ref{fig:sigma_ratio}  compares the ratio of evaporation contributions to charge-changing cross sections. The top panel ({a}) shows the ratio of the estimated evaporation contributions to the experimental charge-changing cross section, given by \( (\sigma_\mathrm{cc}^{\text{expt}} - \sigma_\mathrm{cc}) / \sigma_\mathrm{cc}^{\text{expt}} \) on the left vertical axis, while the ratio \( \sigma_\mathrm{cc}^{\text{evap}} / \sigma_\mathrm{cc}^{\text{expt}} \) is displayed on the right vertical axis. The bottom panel ({b}) represents the ratio of evaporation contributions in the abrasion stage, given by \( \sigma_\mathrm{cc}^{\text{evap}} / \sigma_\mathrm{cc} \), which illustrates the influence of abrasion-stage calculations on the evaporation contributions. The plot compares these ratios for the projectiles \( ^{12} \)C, \( ^{14} \)N, \( ^{16} \)O, and \( ^{20} \)Ne interacting with three different elemental targets at a beam energy of approximately 290 MeV/nucleon.    

The data in Table~\ref{tab:proton_evaporation} demonstrate that reconciling theoretical proton evaporation contributions ($\sigma_\mathrm{cc}^\mathrm{evap}$) with the estimation from experimental data (\(\sigma_\mathrm{cc}^{\text{expt}} - \sigma_\mathrm{cc}\)) requires lowering the parameter \(E_{\text{max}}\) (to \(\sim 15~\text{MeV}\)) for targets with a higher mass number and atomic number, \textit{e.g.}, Al, Cu, compared to lighter ones like C. This adjustment arises from target-dependent differences in initial neutron removal. Heavier targets induce 2--3 neutron removals during the projectile-target interaction in the abrasion stage, creating a neutron-deficient prefragment with an elevated proton-to-neutron ratio. 
Simultaneously, the neutron-deficient prefragment exhibits reduced binding energy, favoring competing decay pathways—particularly neutron re-emission—which deplete excitation energy before protons can evaporate. Since neutron evaporation is not hindered by the Coulomb barrier, it becomes the dominant decay mode, effectively shifting the available excitation spectrum away from high-energy proton emission.  

The parameter \(E_{\text{max}}\), which defines the upper energy limit for proton emission in statistical decay models, must therefore be reduced for Al and Cu targets to reflect the redistribution of excitation energy and the increased dominance of neutron evaporation. Importantly, the target’s role is limited to the initial neutron removal; subsequent evaporation occurs solely from the prefragment, independent of the target.

This suggests that proton evaporation becomes less significant as the mass of the target nucleus increases. In such cases, proton or charge particle removal from the projectile primarily occurs during the abrasion stage or direct fragmentation processes, where the nucleus breaks apart with less contribution from proton evaporation. The trend is more clearly illustrated in Fig.~\ref{fig:sigma_ratio}(b), which indicates that the ratio of evaporation contributions is greater with the carbon target in comparison to the other targets.

Fig.~\ref{fig:sigma_ratio} also reveals distinct trends among the projectiles, with $^{14}$N consistently exhibiting the highest ratio for all target elements. The evaporation of neutrons and protons from the prefragment is governed by different underlying factors.
Neutron emission is primarily determined by the neutron separation energy ($S_n$) and is favored in  neutron-rich nuclei at low excitation energies ($E_\text{ex} \sim S_n$), where the absence of a Coulomb barrier allows evaporation even with limited energy. In contrast, proton evaporation requires overcoming both the proton separation energy ($S_p$) and the Coulomb barrier ($V_c$), making it viable only at high excitation energies ($E_\text{ex} \gg S_p + V_c$), particularly in proton-rich nuclei where excess protons enhance the likelihood of emission. When the initial abrasion stage preferentially removes neutrons from the projectile, the resulting prefragment becomes proton rich, further increasing the probability of proton evaporation. In such cases, the higher proton-to-neutron ratio makes proton emission more favorable compared to neutron emission, even at moderate excitation energies. In neutron rich nuclei, neutron evaporation remains dominant due to lower energy thresholds and higher nuclear level densities at low excitation energy. In neutron-deficient prefragments proton evaporation can become the preferred mode of deexcitation, often reaching near-unity probability as the Coulomb barrier effect diminishes. 
The analysis as shown in Fig.~\ref{fig:proton_evap} highlights the critical role of the separation energy difference, \( \Delta{S_{np}} = S_{1n} - S_{1p} \), in governing charged particle evaporation following neutron removal during abrasion for carbon target. Higher \( \Delta{S_{np}} \) values, such as for \( ^{13}\text{N} \) (\( \Delta S_{np}=18.12 \, \text{MeV} \)), strongly favor proton emission due to the significantly higher energy required for additional neutron removal. Two neutron abrasion further amplifies \( \Delta{S_{np}} \) in neutron-deficient prefragments, {\it e.g.}, $^{14}$O 
($\Delta S_{np}=18.56$~MeV), driven by symmetry energy and Coulomb effects~\cite{Symmetry2019}. These trends underscore \( \Delta{S_{np}} \) as a key predictor of evaporation pathways in systems with fewer neutrons relative to protons.

The overall dominance of proton evaporation from the prefragment $^{13}$N can be better understood by using a simplified discussion on the ratio of proton to neutron decay width~\cite{benlliure1998,de2020},
\begin{equation}
\frac{\Gamma_p}{\Gamma_n} \approx \frac{\exp \left[ 2 \sqrt{a (E_\mathrm{ex} -  S_p  - V_c-\delta_{ \rm eff}}) \right]}{\exp \left[ 2 \sqrt{a (E_\mathrm{ex} -  S_n -\delta_{ \rm eff}}) \right]},
\end{equation}
with $\delta_{ \rm eff}$ as the pairing energy correction.
The level density parameter is given by
\begin{equation}
a = \tilde{a} \left[ 1 + \frac{\Delta S}{U} \left( 1 + \exp(-\gamma U) \right) \right],
\end{equation}
as suggested by Ignatyuk~\cite{ignatyuk1975}, where $\tilde{a}$ is the asymptotic level density parameter, this parameter is defined  
$\tilde{a} = 0.073A + 0.095A^{2/3}.$ $\Delta S$ as the shell correction obtained from the difference between the experimental and the liquid drop model masses, $U$ is the effective excitation energy and $\gamma$ as the shell damping factor. The \textsuperscript{13}N prefragment, formed from \textsuperscript{14}N after one-neutron removal in the abrasion stage, is characterized by a high level density, a low proton separation energy ($S_p$), a high neutron separation energy ($S_n$), and a low Coulomb barrier. These factors contribute to a significantly high decay width ratio (${\Gamma_p/\Gamma_n}$), as illustrated in Fig.~\ref{fig:gamma_ratio}. This strongly favours proton evaporation, making it a dominant contributor to the charge-changing cross sections through proton emission. \textsuperscript{11}C, the prefragment of \textsuperscript{12}C after $1n$ removal in the abrasion stage, has the lowest decay width ratio (${\Gamma_p/\Gamma_n}$) and consequently exhibits the smallest contribution to ($\sigma_{\rm cc}^{\rm evap}$) in Fig.~\ref{fig:sigma_ratio}, making it a limited contributor to
charge-changing cross sections through proton emission.

\begin{figure}
    \centering  \includegraphics[width=0.7\columnwidth]{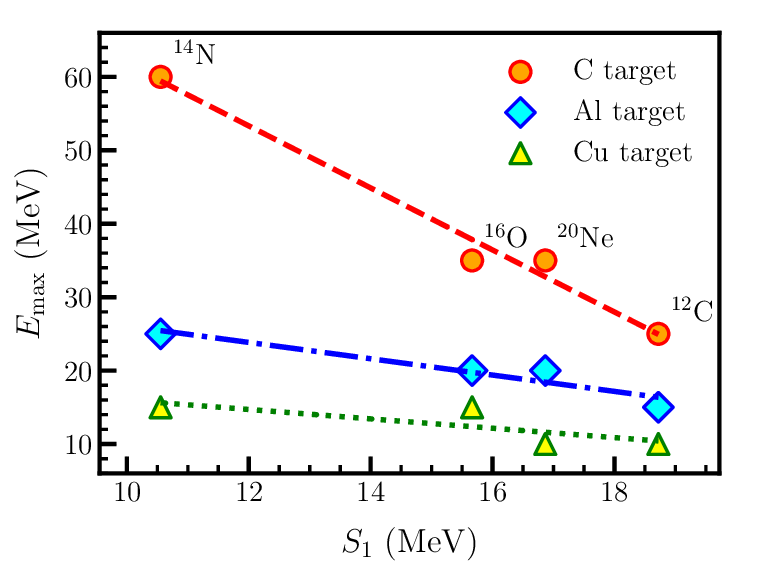}
   \caption{Correlation between one neutron separation energy ($S_{1}$) from the projectile and the maximum excitation energy parameter ($E_{\text{max}}$) for different target materials (C, Al, Cu). The scatter points represent individual data for the projectiles $^{12}$C, $^{14}$N, $^{16}$O, and $^{20}$Ne, while the fitted lines illustrate the underlying trends for each target.}
   \label{fig:Emax_vs_sn}
\end{figure}

The one neutron separation energy (\( S_1 \)) of the projectile further influences the residual excitation energy available for evaporation.  High \( S_1 \) in neutron-deficient projectiles depletes excitation energy during neutron removal, reducing ablation, whereas low \( S_1 \) in neutron-rich systems preserves energy, enhancing ablation. Thus, projectiles with lower \( S_1 \) set a higher \( E_\mathrm{max} \), as evidenced by the linear \( E_\mathrm{max} \)-\( S_1 \) correlation in Fig.~\ref{fig:Emax_vs_sn}, which holds across all targets. 

These effects highlight the necessity of dynamically adjusting the maximum excitation energy parameter (\( E_{\text{max}} \)) in the evaporation model. Incorporating projectile specific properties—including neutron separation energy, neutron orbital occupancy, and beam energy—is crucial for accurately reproducing the observed dominance of abrasion (direct nucleon removal) in  heavier target systems and ablation (evaporation) in lighter or neutron rich projectiles.

\section{\label{sec:level4}Summary and conclusions}

In this study, we calculated nuclear charge-changing cross sections for projectile nuclei, including $^{12}$C, $^{14}$N, $^{16}$O, and $^{20}$Ne, interacting with elemental targets such as carbon, aluminum, and copper at a beam energy of approximately 290 MeV/nucleon. We employed the abrasion-ablation model, which describes the fragmentation process in two stages -- an initial abrasion phase, where nucleons are removed due to high-energy collisions, followed by an ablation phase, where the excited residual nucleus undergoes particle evaporation. Our calculations utilized the finite-range optical limit approximation within the Glauber model to describe the abrasion step of nuclear fragmentation, incorporating realistic nuclear density distributions. Additionally, we analyzed the ablation process by evaluating the evaporation contribution to charge-changing cross sections through a statistical decay model and by subtracting the abrasion component from the experimental data. We further examined how factors such as excitation energy distribution, particle decay widths, and separation energies influence charged particle evaporation and its impact on charge-changing cross sections. Our findings indicate that for heavier targets, the maximum excitation energy parameter ($E_{\mathrm{max}}$) tends to be lower when fitted to experimental data. Specifically, for a carbon target, we determined a higher $E_{\mathrm{max}}$ value around $35 \pm 10$ MeV, with the exception of $^{14}$N, which exhibited an anomalous fitting around 60 MeV. In contrast, for aluminum and copper targets, the excitation energy parameter was consistently lower, around $15 \pm 10$ MeV. Furthermore, our analysis revealed an inverse linear correlation between $E_{\mathrm{max}}$ and the neutron separation energy of the projectile, highlighting the importance of projectile structure in determining the fragmentation process.

The ablation stage plays a crucial role in shaping the charge-changing cross-section measurements by influencing the final distribution of nuclear fragments. Our study underscores the necessity of precise modeling of the ablation process to improve the accuracy of theoretical predictions and experimental interpretations. In space, charge-changing cross sections are used to predict cosmic ray interactions with shielding and human tissues, aiding astronaut safety. In cancer therapy, they influence ion beam fragmentation, impacting dose delivery and treatment precision. Understanding these processes thus enhance space radiation protection, hadron therapy accuracy, and nuclear reactor safety. Moreover, our findings contribute to broader applications in astrophysics and cosmic ray propagation studies, where accurate modeling of nuclear fragmentation is essential.

Future work could explore refinements to the excitation energy parameterization by incorporating microscopic nuclear structure effects and extending calculations to a wider range of projectile-target combinations. Additionally, further experimental validation of the ablation model could help refine predictions, ultimately enhancing our understanding of nuclear fragmentation mechanisms.

\begin{acknowledgments}
The Indo-Japan Cooperative Science Programme (IJCSP)-2023 under the DST-JSPS bilateral programs JPJSBP120247715 and
DST/INT/JSPS/P-393/2024(G). [S.D] also acknowledges financial support from the Ministry of Education (MoE), India, through a 
doctoral fellowship.
\end{acknowledgments}

\bibliography{apssamp}

\end{document}